\documentclass[12pt,aps,prd,preprint,tightenlines,
  showpacs,nofootinbib]{revtex4}
\newcommand{\PRE}[1]{{#1}} 

\usepackage{bm}
\usepackage{epsfig}

\newcommand{\kev}{\text{keV}}
\newcommand{\mev}{\text{MeV}}
\newcommand{\gev}{\text{GeV}}
\newcommand{\tev}{\text{TeV}}

\newcommand{\cm}{\text{cm}}

\newcommand{\km}{\text{km}}
\newcommand{\g}{\text{g}}
\newcommand{\s}{\text{s}}

\newcommand{\Mpc}{\text{Mpc}}

\newcommand{\pc}{\text{pc}}
\newcommand{\etal}{{\em et al.}}

\newcommand{\eqref}[1]{Eq.~(\ref{#1})}
\newcommand{\eqsref}[2]{Eqs.~(\ref{#1}) and (\ref{#2})}
\newcommand{\secref}[1]{Sec.~\ref{sec:#1}}
\newcommand{\secsref}[2]{Secs.~\ref{sec:#1} and \ref{sec:#2}}
\newcommand{\appref}[1]{App.~\ref{sec:#1}}
\newcommand{\figref}[1]{Fig.~\ref{fig:#1}}

\newcommand{\mgravitino}{m_{\gravitino}}

\newcommand{\gravitino}{\tilde{G}}

\newcommand{\be}{\begin{equation}}
\newcommand{\ee}{\end{equation}}

\newcommand{\TRH}{T_{\text{RH}}}
\newcommand{\xiRH}{\xi_{\text{RH}}}
\newcommand{\ThRH}{T^h_{\text{RH}}}

\newcommand{\uone}{\text{U(1)}_{\text{EM}}}
\newcommand{\tantheta}{\tan\theta_W^h}
\newcommand{\Mcut}{M_{\text{cut}}}
\newcommand{\kcut}{k_{\text{cut}}}

\begin{document}

\preprint{UCI-TR-2009-06}

\title{
\PRE{\vspace*{1.5in}} Hidden Charged Dark Matter
\PRE{\vspace*{0.3in}} }

\author{Jonathan L.~Feng, Manoj Kaplinghat, Huitzu Tu, and Hai-Bo Yu}
\affiliation{Department of Physics and Astronomy, University of
California, Irvine, California 92697, USA \PRE{\vspace*{.5in}} }


\begin{abstract}
\PRE{\vspace*{.3in}} Can dark matter be stabilized by charge
conservation, just as the electron is in the standard model?  We
examine the possibility that dark matter is hidden, that is, neutral
under all standard model gauge interactions, but charged under an
exact U(1) gauge symmetry of the hidden sector.  Such candidates are
predicted in WIMPless models, supersymmetric models in which hidden
dark matter has the desired thermal relic density for a wide range of
masses.  Hidden charged dark matter has many novel properties not
shared by neutral dark matter: (1) bound state formation and
Sommerfeld-enhanced annihilation after chemical freeze out may reduce
its relic density, (2) similar effects greatly enhance dark matter
annihilation in protohalos at redshifts of $z \sim 30$, (3) Compton
scattering off hidden photons delays kinetic decoupling, suppressing
small scale structure, and (4) Rutherford scattering makes such dark
matter self-interacting and collisional, potentially impacting
properties of the Bullet Cluster and the observed morphology of
galactic halos.  We analyze all of these effects in a WIMPless model
in which the hidden sector is a simplified version of the minimal
supersymmetric standard model and the dark matter is a hidden sector
stau.  We find that charged hidden dark matter is viable and
consistent with the correct relic density for reasonable model
parameters and dark matter masses in the range $1~\gev \alt m_X \alt
10~\tev$.  At the same time, in the preferred range of parameters,
this model predicts cores in the dark matter halos of small galaxies
and other halo properties that may be within the reach of future
observations. These models therefore provide a viable and
well-motivated framework for collisional dark matter with Sommerfeld
enhancement, with novel implications for astrophysics and dark matter
searches.
\end{abstract}


\maketitle

\section{Introduction}
\label{sec:intro}

Dark matter makes up most of the matter in the Universe, but its
identity is unknown. So far, dark matter has been observed only
through its gravitational interactions.  A logical possibility, then,
is that dark matter is hidden, that is, neutral under all standard
model (SM) gauge interactions.  This possibility has been explored for
many years~\cite{Kobsarev:1966,Blinnikov:1982eh,Hodges:1993yb,%
Berezhiani:1995am,Mohapatra:2000qx,Berezhiani:2000gw,Mohapatra:2001sx,%
Ignatiev:2003js,Foot:2003iv,Berezhiani:2003wj,Foot:2004wz} and brings
with it a great deal of model building freedom.  Unfortunately, this
freedom is generically accompanied not only by a loss of predictivity,
but also by the loss of appealing features common to other dark matter
candidates, such as the strong connections to central problems in
particle physics, such as the gauge hierarchy and flavor problems, and
the naturally correct thermal relic density of weakly-interacting
massive particles (WIMPs).

In this work, we consider the possibility that hidden dark matter is
charged under an {\em exact}\ U(1) gauge symmetry in the hidden
sector. We assume the hidden sector is truly hidden, with no
``connector'' particles charged under both SM and hidden gauge groups.
Although many of our results are relevant for any charged hidden dark
matter, including cases with connectors or massive hidden photons, we
are particularly motivated by the framework of WIMPless dark
matter~\cite{Feng:2008ya,Feng:2008dz,Feng:2008qn,Feng:2008mu,%
Kumar:2009af,McKeen:2009rm}, which ameliorates some of the problems of
hidden dark matter noted above.  (For other recent discussions of
hidden dark matter charged under exact gauge symmetries, see
Refs.~\cite{Ackerman:2008gi,Kaloper:2009nc,Dai:2009hx}.) In this
framework, dark matter is hidden, but additional structure implies
that it nevertheless naturally has the correct thermal relic density.
In the examples studied in Refs.~\cite{Feng:2008ya,Feng:2008dz,%
Feng:2008qn,Feng:2008mu,Kumar:2009af,McKeen:2009rm}, the dark matter
is in a hidden sector of supersymmetric models with gauge-mediated
supersymmetry breaking (GMSB).  Assuming that supersymmetry is broken
in a single sector, the hidden sector mass scale $m_X$ and gauge
couplings $g_X$ are related to the observable sector weak scale $m_W$
and gauge coupling $g_W$ by
\begin{equation}
\frac{m_X}{g_X^2} \sim \frac{m_W}{g_W^2} \ .
\end{equation}
This implies that the thermal relic density of WIMPless dark matter,
$\Omega_X \sim \langle \sigma v \rangle ^{-1} \sim m_X^2/ g_X^4 \sim
m_W^2 / g_W^4 \sim \Omega_{\text{WIMP}}$, and so is naturally of the
desired magnitude, even though $m_X$ need not be at the weak scale.
The WIMPless framework also has other virtues, including that fact
that it is naturally consistent with an elegant solution to the new
physics flavor problem, since the gravitino mass, and with it all
dangerous flavor-violating gravity-mediated effects, satisfies
$\mgravitino \ll m_W$.  This contrasts with the case of neutralino
dark matter, where stability of the neutralino $\chi$ requires
$\mgravitino > m_{\chi} \sim m_W$.  Other GMSB dark matter candidates
have been proposed~\cite{Han:1997wn,Baltz:2001rq,Ibe:2006rc,%
Feng:2008zza,Hooper:2008im}.  However, it is notable that the relation
$m_X \propto g_X^2$ results directly from the desire to generate a
flavor-blind superpartner spectrum.  WIMPless dark matter therefore
not only dissolves the tension between dark matter and the flavor
problem in supersymmetry, but it exploits an elegant solution of the
supersymmetric flavor problem to predict the correct thermal relic
density.

Assuming $\mgravitino < m_X$, however, how is the stability of the
hidden sector candidate maintained?  Although there are many
possibilities in the WIMPless framework~\cite{Feng:2008ya}, an elegant
possibility is that the dark matter is charged under an exact gauge
symmetry in the hidden sector.  The dark matter candidate may then be
stable for the same reason that the electron is stable --- its decay
is prevented by charge conservation.  We are therefore led rather
straightforwardly to the possibility of charged dark matter in a
hidden sector.  In the following sections, we will consider, as a
concrete example, a minimal supersymmetric standard model (MSSM)-like
hidden sector with ${\cal O}(1)$ Yukawa couplings.  The massless
hidden particles --- the hidden photon, gluon, and neutrinos --- are
then all neutral under hidden $\uone$.  Any charged particle in the
hidden sector, for example, the hidden $W$, chargino, stau, tau, or
charged Higgs boson, would then be stable, provided it was the
lightest $m_X$ particle.  A particularly plausible choice is the
lighter stau $\tilde{\tau}$, whose mass receives smaller SUSY-breaking
contributions than other hidden superparticles.

This possibility of charged hidden dark matter requires a rethinking
of many issues in cosmology, as its properties differ markedly from
the more conventional possibility of neutral dark matter.  In this
work, we analyze these issues comprehensively to establish the
viability of the scenario and determine what ranges of parameters are
allowed.  Because the dark matter is charged, it may form bound states
and then annihilate in the early Universe.  Dark matter annihilation
is also enhanced by the Sommerfeld effect~\cite{Sommerfeld:1931}.  The
formation of bound states and the Sommerfeld effect are possible also
in the standard model; for example, bound state formation impacts
$e^+e^-$ annihilation to 511 keV photons at late times when the $e^+$
and $e^-$ are non-relativistic.  In the present case, both effects may
reduce the thermal relic density, potentially destroying the nice
relic density property of WIMPless dark matter.  After chemical freeze
out, the dark matter's charge also plays an important role in kinetic
decoupling, since charged dark matter may be far more efficiently
coupled to the thermal bath of hidden photons through Compton
scattering than neutral dark matter.  This impacts the matter power
spectrum, requiring a modification of the standard WIMP
analysis~\cite{Chen:2001jz,Berezinsky:2003vn,Green:2003un}.  This also
changes the expected minimum mass of dark matter halos. It is
important to check the minimum halo mass because the Sommerfeld
enhancement charged dark matter annihilation is most effective in the
least massive halos where relative velocities are the smallest, and
this effect has been noted to provide a stringent constraint on the
possibility of charged dark matter~\cite{Kamionkowski:2008gj}.
Finally, the existence of a long range force between dark matter
particles implies dark matter is collisional. It is therefore also
critical to investigate whether this scenario is consistent with
constraints on self-interacting dark matter from the Bullet
Cluster~\cite{Randall:2007ph}, observed ellipticity of dark matter
halos~\cite{MiraldaEscude:2000qt} and bounds from considerations of
galactic dynamics~\cite{Ackerman:2008gi}.

In the end, we will find that all of these constraints are satisfied
for reasonable model parameters, and are consistent with the correct
relic density for dark matter masses in the range $1~\gev \alt m_X
\alt 10~\tev$.  These dark matter candidates are therefore viable,
sharing many of the theoretical virtues of WIMPs, but predicting
drastically different features for astrophysics.  In its minimal form,
this model also predicts no direct detection, indirect detection, and
collider signals.  If desired, however, with the addition of connector
sectors, the charged hidden dark matter model we consider here may
also have implications for direct searches, indirect searches, and
particle colliders.  Some of these implications have been considered
previously~\cite{Feng:2008ya,Feng:2008dz,Feng:2008qn,Feng:2008mu,%
Kumar:2009af,McKeen:2009rm}, and others are
possible~\cite{Pospelov:2007mp,Jaeckel:2008fi,%
Krolikowski:2008pq,ArkaniHamed:2008qn,Gardner:2008yn,Chun:2008by,%
MarchRussell:2008tu,Baumgart:2009tn,Shepherd:2009sa,Cheung:2009qd,%
Katz:2009qq,Batell:2009yf,Essig:2009nc,Gopalakrishna:2009yz,%
Morrissey:2009ur,Bi:2009am}.  We will not expand on these
possibilities here, other than to note that our results show that an
exact hidden $\uone$ is possible, leading to a simple framework in
which annihilation rates are enhanced by the pure Sommerfeld effect.

This paper is organized as follows.  In \secref{model}, we briefly
review our hidden sector model and its parameters, and discuss its
thermal relic abundance.  In \secref{kinetic}, we study the kinetic
decoupling of the hidden sector dark matter, and in \secref{matter} we
present the resulting matter power spectrum and derive the minimum
mass of dark matter halos. Details of this derivation are given in
\secref{app}. The impact of bound states and enhanced annihilation on
relic densities is evaluated in \secref{boundstate}, and the
constraint from annihilation in protohalos is discussed in
\secref{protohalo}.  In \secref{halo}, we study the effect of the long
range force between dark matter particles on galaxy shapes and derive
bounds on the parameter space. In \secref{bullet} we verify that our
model is consistent with bounds on self-interactions, such as those
from merging clusters.  We summarize our conclusions in
\secref{conclusions}.  We note that this work is necessarily rather
wide-ranging; readers interested in only a subset of the topics may
find \secref{conclusions} a useful guide to the relevant conclusions
and figures.

\section{Hidden Sector: Particles, Parameters, and Relic Density}
\label{sec:model}

In this work, following Ref.~\cite{Feng:2008mu}, we consider a hidden
sector that is MSSM-like, with gauge group $\text{SU(3)} \times
\text{SU(2)}_L \times \text{U(1)}_Y$, but with only one generation of
fermions and ${\cal O}(1)$ Yukawa couplings.  It is important that the
model be chiral, so that hidden electroweak symmetry breaking sets the
scale for the masses and the naturally correct thermal relic density
is realized.  Of course, the model should also be anomaly-free.  The
one-generation MSSM is a relatively simple model that satisfies these
criteria, and has the advantage that one can apply much of the
intuition built up from studies of the usual MSSM.  Many other choices
are possible, however, and would also be worth further study.
Assuming electroweak symmetry breaking that leaves $\text{SU(3)}
\times \uone$ preserved in the hidden sector, the particle masses in
the hidden sector are
\begin{eqnarray}
\text{Hidden~weak~scale~:} && W^h, Z^h, h^h, \tilde{g}^h,
\chi^{\pm \, h}, \chi^{0 \, h}, t^h, b^h, \tau^h, \tilde{t}^h,
\tilde{b}^h, \tilde{\tau}^h, \tilde{\nu}_{\tau}^h \nonumber \\
\sim 0: && \gamma^h, g^h, \nu_{\tau}^h, \tilde{G}  \ ,
\end{eqnarray}
where we have used third-generation notation for the hidden
(s)quarks and (s)leptons to remind us that they have ${\cal O}(1)$
Yukawa couplings.  {}From this spectrum, a natural possibility is
that the lightest weak scale particle is charged, and therefore
stabilized by $\uone$ charge conservation in the hidden sector. As
noted in \secref{intro}, there are many possible charged dark matter
candidates in the hidden sector.  To be concrete, we will focus here
on the right-handed stau, which we denote $\tilde{\tau}^h$ from now
on.

Despite the complexity of the hidden sector, with a few mild
assumptions, there are only a few parameters that are relevant for
this study.  These are
\begin{equation}
m_X, \ m_{Z^h}, \ \alpha_X, \ \tantheta, \ \xiRH \equiv
\frac{\ThRH}{\TRH} \ ,
\label{parameters}
\end{equation}
where $m_X = m_{\tilde{\tau}^h}$ is the dark matter particle's mass,
$m_{Z^h}$ is the mass of the hidden $Z$ boson, $\alpha_X = (e^h)^2 / 4
\pi$ is the fine structure constant of the hidden $\uone$,
$\theta^h_W$ is the hidden sector's weak mixing angle, and $\xiRH$ is
the ratio of hidden to observable sector temperatures at the time of
reheating.  The parameter $\xiRH$ need not be 1; for example, the
inflaton may couple differently to the observable and hidden sectors,
leading to a temperature
asymmetry~\cite{Hodges:1993yb,Berezhiani:1995am}.  For a
one-generation MSSM to satisfy Big Bang nucleosynthesis constraints on
massless degrees of freedom, one requires $\xiRH \alt
0.8$~\cite{Feng:2008mu}.  We will consider values in the range $0.1
\le \xiRH \le 0.8$.  In the hidden sector, the $\gamma^h$ and $\nu^h$
temperatures are always identical, and we denote the common
temperature by $T^h$.  In the visible sector, the $\gamma$ and $\nu$
temperatures diverge eventually, of course; we denote let $T$ denote
the $\gamma$ (cosmic microwave background (CMB)) temperature.  We do
not assume hidden sector gauge unification, although we will present
some of our results for the grand unified value $\tantheta =
\sqrt{3/5} \simeq 0.8$ as an interesting example.  We assume that the
hidden sector's strong coupling $g_s$ is small enough that
hadronization effects are negligible in the early Universe.

The hidden dark matter's relic abundance is determined by the
parameters of \eqref{parameters}.  The dark matter annihilates through
$\tilde{\tau}^{h\, +} \tilde{\tau}^{h\, -} \to \gamma^h \gamma^h,
\nu^h \bar{\nu}^h$.  If the $\tilde{\tau}^h$ is dominantly
right-handed, its thermally-averaged annihilation cross section
is~\cite{Feng:2008mu}
\begin{equation}
\label{partial_wave}
   \langle \sigma_A v \rangle (T^h) = \frac{(4\pi \alpha_X)^2}{m^2_X}
   \left[ a_0 +
   a_1 \frac{T^h}{m_X} + {\cal O} \left( \frac{T^h}{m_X}\right)^2
\right]  ,
\end{equation}
where
\begin{eqnarray}
\label{partial_wave2}
   a_0 &=& \left[\frac{1}{8 \pi} + \frac{1}{4 \pi}\,
   \left(1 - \frac{m^2_{Z^h}}{4 m^2_X} \right)\,
   \tan^2 \theta^h_W \right]\, \nonumber \\
   a_1 &=& - \frac{1}{4 \pi} - \frac{1}{2 \pi}\,
   \tan^2 \theta^h_W + \frac{1}{8 \pi}\, \frac{1}{\cos^4 \theta^h_W\,
   \left[\left(-4 + \frac{m^2_{Z^h}}{m^2_X} \right)^2 + \frac{m^2_{Z^h}
   \Gamma^2_{Z^h}}{m^4_X} \right]} \ .
\end{eqnarray}

To avoid presenting results for the special case in which the
annihilation cross section is resonantly enhanced by the $Z^h$ pole,
we fix $m_{Z^h} = 1.5 m_X$.  The dependence of $\langle \sigma_A v
\rangle$ on $\Gamma_{Z^h}$ is therefore negligible.  In the numerical
studies below, we will consider weak mixing angles in the range
$\sqrt{3/5} \le \tantheta \le 10$.  If the hidden sector has custodial
symmetry, $m_{W^h}=m_{Z^h}\cos\theta^h_W$, then $\tantheta > 1.11$
implies $m_{W^h} < m_{\tilde{\tau}^h}$.  Although $W^h$ would also be
a perfectly good charged hidden dark matter candidate, we will instead
assume custodial symmetry is broken, for example, by triplet Higgs
bosons, so that $\tilde{\tau}^h$ remains the dark matter candidate for
all values of $\tantheta$.

\begin{figure}[tb]
\begin{center}
\includegraphics*[width=12cm,clip=]{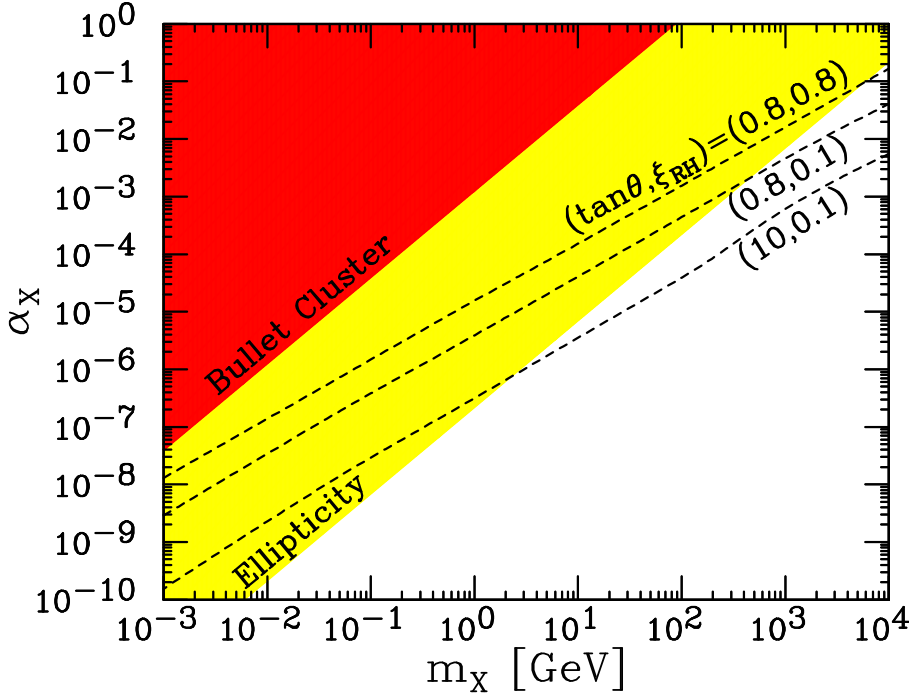}
\end{center}
\vspace*{-.2in}
\caption{Allowed regions in $(m_X, \alpha_X)$ plane, where $m_X$ is
the mass of the dark matter charged under the unbroken hidden sector
$\uone$ with fine-structure constant $\alpha_X$.  Contours for fixed
dark matter cosmological relic density consistent with WMAP results,
$\Omega_X h^2 = 0.11$, are shown for $(\tantheta, \xiRH) =
(\sqrt{3/5}, 0.8)$, $(\sqrt{3/5},0.1)$, $(10, 0.1)$ (dashed), from top
to bottom, as indicated.  The shaded regions are disfavored by
constraints from the Bullet Cluster observations on self-interactions
(dark red) and the observed ellipticity of galactic dark matter halos
(light yellow).  The Bullet Cluster and ellipticity constraints are
derived in \secsref{bullet}{halo}, respectively.
\label{fig:alpha_constraint}}
\end{figure}

With these assumptions, contours of constant $\Omega_X h^2$ in the
$(m_X, \alpha_X)$ plane are given in \figref{alpha_constraint}.
(Constraints from observations are also shown in
\figref{alpha_constraint}; these will be discussed in
\secsref{halo}{bullet}.) As evident from \figref{alpha_constraint} and
\eqref{partial_wave2}, there is a strong dependence on $\tantheta$:
for fixed $\alpha_X$, the neutrino cross section is enhanced by
$1/\cos^4\theta_W^h$ for large $\tantheta$.  This dependence is very
important. In high energy processes, such as thermal freeze out, both
EM and weak processes are effective.  In low energy processes, such as
those occurring at present, only EM interactions mediated by the
massless $\gamma^h$ are important.  If there were no hidden weak
interactions, for fixed $m_X$ and $\xiRH$, $\Omega_X h^2 \simeq 0.11$
would fix $\alpha_X$, and so the relic density would fix the strength
of dark matter self-interactions now.  In the more general case we are
considering, however, one can keep $\Omega_X$ constant for arbitrarily
small $\alpha_X$, provided $\tantheta$ is large enough.  As noted
above, we will consider $\tantheta$ as large as 10; larger values are,
of course, also possible.

\section{Kinetic Decoupling of Dark Matter}
\label{sec:kinetic}

In the standard cosmology, structure formation is hierarchical, that
is, small objects form first and progressively the larger objects form
via mergers and accretion. The mass of the smallest dark matter halo
is determined by the dark matter particle's mass and the temperature
at which it kinetically decouples from the other light particle
species. For a typical WIMP with electroweak scale mass, the kinetic
decoupling temperature ranges roughly from $10~\mev$ to $1~\gev$,
which leads to the formation of dark matter halos with masses as small
as $10^{-12} M_{\odot}$ to $10^{-4} M_{\odot}$~\cite{Chen:2001jz,%
Berezinsky:2003vn,Green:2003un,Green:2005fa,Loeb:2005pm,%
Profumo:2006bv,Bertschinger:2006nq,Bringmann:2006mu}.  For other dark
matter particles, such as weak-scale gravitinos or MeV dark matter,
the minimum mass dark matter halo could be as large as the smallest
dwarf
galaxies~\cite{Kaplinghat:2005sy,Cembranos:2005us,Hooper:2007tu}.
Some regions of the parameter space of these models are excluded
because the predicted minimum mass halo is in conflict with
observations.

In this section, we analyze the kinetic decoupling of hidden charged
dark matter.  One notable difference between the WIMP and hidden
charged dark matter is that the charged dark matter interacts not only
through weak interactions, but also through EM interactions.  For the
case of $\tilde{\tau}^h$ dark matter, this implies that the dark
matter remains in kinetic contact not only through the weak process
$\tilde{\tau}^h \nu^h \leftrightarrow \tilde{\tau}^h\nu^h$, but also
through the Compton scattering process $\tilde{\tau}^h \gamma^h
\leftrightarrow \tilde{\tau}^h \gamma^h$.  As we will see, at low
temperatures, the thermally-averaged weak cross section is suppressed
by $T^{h\, 2}/m_X^2$, but this suppression is absent for Compton
scattering, creating a large, qualitative difference between this case
and the canonical WIMP scenario.  Note also that, in principle, in the
case of charged dark matter, bound state formation also impacts
kinetic decoupling.  As we will see in \secref{boundstate}, however,
very few staus actually bind, and so this effect is not significant
and may be neglected in our analysis.

We follow Refs.~\cite{Bertschinger:2006nq,Bringmann:2006mu} to
determine the temperature of kinetic decoupling for the dark matter
particle.  
In the hidden sector, the Boltzmann equation governing the
evolution of the dark matter particle's phase space distribution is
\begin{equation}
  \frac{df(\vec{p})}{dt} = \Gamma(T^h)(T^h m_X
   \triangle_{\vec{p}}+\vec{p}\cdot\nabla_{\vec{p}}+3)f(\vec{p})\ ,
\label{eq:kinetic_coupling}
\end{equation}
where
\begin{equation}
  \Gamma (T^h) = \sum_n \frac{g_0}{6(2\pi)^3} m_X c_n N^{\pm}_{n+3}
\left( \frac{T^h}{m_X} \right)^{n+4}
\label{Gamma}
\end{equation}
is the momentum transfer rate.  In \eqref{Gamma}, $g_0$ is the number
of degrees of freedom of the scattering particle ($g_0 = 2$ for both
$\gamma^h$ and $\nu^h$), the $N^{\pm}_{n+3}$ are constants defined in
Ref.~\cite{Bringmann:2006mu}, and the $c_n$ are determined by
parameterizing the collision amplitudes, evaluated at $t=0$ and $s =
m_X^2 + 2 m_X T^h$, as
\begin{equation}
\left| {\cal M} \right|^2_{t=0, s=m_X^2 + 2 m_X T^h} \equiv
c_n \left( \frac{T^h}{m_X} \right)^n +
{\cal O} \left( \frac{T^h}{m_X} \right)^{n+1} \ .
\label{expansion}
\end{equation}

In the present case, with the help of the CalcHEP
program~\cite{Pukhov:1999gg,Pukhov:2004ca}, we find that the squared
amplitudes of the relevant processes are
\begin{eqnarray}
  \left| {\cal M}(\tilde{\tau}^h\gamma^h \leftrightarrow
\tilde{\tau}^h\gamma^h) \right|^2
  &=& \frac{64\pi^2\alpha^2_X
    \left[(m^2_X-s)^4+2(m^2_X-s)^2st+(m^4_X+s^2)t^2 \right]}
  {(m^2_X-s)^2(-m^2_X+s+t)^2} \\
  \left| {\cal M}(\tilde{\tau}^h \nu^h \leftrightarrow
\tilde{\tau}^h \nu^h) \right|^2
  &=& \frac{16 \pi^2 \alpha^2_X\,[(m_X^2-s)^2+st]}
  {\cos^4\theta^h_W\, (m^2_Z-t)^2} \ ,
\end{eqnarray}
and so
\begin{eqnarray}
  \left| {\cal M}(\tilde{\tau}^h\gamma^h \leftrightarrow
\tilde{\tau}^h\gamma^h) \right|^2_{t=0, s = m_X^2 + 2 m_X T^h}
&=& 64\pi^2\alpha^2_X  \label{Mgg} \\
   \left|{\cal M}(\tilde{\tau}^h\nu^h \leftrightarrow
\tilde{\tau}^h\nu^h) \right|^2_{t=0, s = m_X^2 + 2 m_X T^h}
   &=&\frac{64\pi^2\alpha^2_Xm^4_X}{\cos^4\theta^h_W
   m^4_Z}\left(\frac{T^h}{m_X}\right)^2 \ . \label{Mnunu}
\end{eqnarray}
Casting these results in the form of \eqref{expansion}, we find
\begin{equation}
  \Gamma (T^h) = \frac{32 \pi^3 \alpha^2_X\, T^{h\, 4}}
  {45\, m^3_X}\, +
   \frac{124 \pi^5 \alpha^2_X\, m_X\, T^{h\, 4}}
  {63 \cos^4 \theta^h_W\, m^4_Z}\, \left(\frac{T^h}{m_X} \right)^2 \ .
\label{rate}
\end{equation}

The inverse of $\Gamma(T^h)$ is the time needed for the
hidden sector photons and neutrinos to transfer momentum $|\vec{p}|
\sim T^h$ to the staus and thereby keep the dark matter in
kinetic equilibrium. During this phase, the temperature of the dark
matter tracks that of the hidden section photons or neutrinos,
depending on the dominant scattering mechanism.  The contribution from
Compton scattering (the first term) does not suffer the
$(T^h/m_X)^2$ suppression typical of weak interactions.
Therefore, for $\tantheta \sim 1$, Compton scattering dominates the
contribution to $\Gamma(T^h)$. However, for $\tantheta=10$, the
$\nu^h$ channel is enhanced dramatically and dominates for $m_X \alt
1~\gev$.

We define the kinetic decoupling temperature as the temperature where
the momentum transfer rate and the Hubble expansion rate become equal,
that is, $\Gamma(T^h_{\text{kd}})=H(T^h_{\text{kd}})$.  (For other
possible definitions, see, for example,
Refs.~\cite{Bertschinger:2006nq,Bringmann:2006mu}.)  Here we adopt the
hidden sector point of view and write the Hubble parameter as
\begin{eqnarray}
  H (T^h)=\sqrt{\frac{4\pi^3}{45} g^{\rm tot}_\ast (T^h)}\,
   \frac{T^{h\, 2}}{M_{\rm pl}}\, ,
\label{eq:hubble}
\end{eqnarray}
where $g^{\rm tot}_\ast (T^h) = g^h_\ast + g_\ast \left( T / T^h
\right)^4$ is the total effective relativistic degrees of freedom, and
$T$ is the visible sector's photon temperature.  We assume there is no
thermal contact between hidden and observable sectors, and so the
visible sector enters the analysis only through the contribution of
its effective degrees of freedom $g_\ast$ to the expansion rate.

\begin{figure}[tb]
\begin{center}
\includegraphics*[width=12cm,clip=]{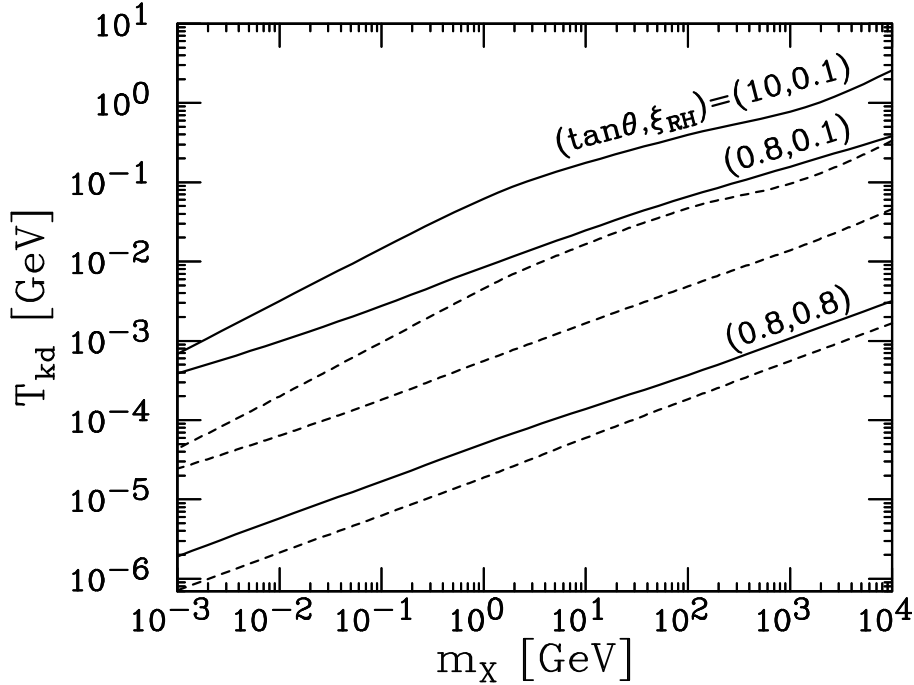}
\end{center}
\vspace*{-.2in}
\caption{Kinetic decoupling temperatures as functions of dark matter
mass $m_X$ for $(\tantheta, \xiRH) = (10, 0.1)$, $(\sqrt{3/5},0.1)$,
$(\sqrt{3/5}, 0.8)$, from top to bottom, as indicated.  For each case
we plot both the hidden sector photon temperature $T^h_{\text{kd}}$
(dashed) and the corresponding visible sector photon (CMB) temperature
$T_{\text{kd}}$ (solid) at the time of kinetic decoupling.}
\label{fig:Tkd}
\end{figure}

In \figref{Tkd} we show the hidden sector kinetic decoupling
temperature $T_{\text{kd}}^h$ and visible sector kinetic decoupling
temperature $T_{\text{kd}}$ as functions of $m_X$ for various
combinations of $(\tantheta, \xiRH)$.  The hidden $\uone$
fine-structure constant $\alpha_X$ is determined by requiring the
correct dark matter relic abundance $\Omega_X h^2 = 0.11$.  To
understand these results, consider first the case $(\tantheta, \xiRH)
= (\sqrt{3/5}, 0.8)$ and $m_X \sim 100~\gev$.  In this case, $g_1 \sim
g_2$ in the hidden sector, the hidden and observable sectors have
comparable temperatures, and the dark matter has a weak-scale mass.
This case is thus similar to the standard WIMP case, except that the
hidden dark matter is charged.  For the WIMP case, one typically finds
$T_{\text{kd}} \sim 10~\mev - 1~\gev$, as noted above, whereas in this
charged hidden dark matter case, we find that $T_{\text{kd}} \sim
0.1~\mev$.  We see that the presence of Compton scattering does in
fact have a large impact, keeping the charged dark matter in kinetic
equilibrium to much lower temperatures.

This effect is moderated for other values of $(\tantheta, \xiRH)$.
For $(\tantheta, \xiRH) = (\sqrt{3/5}, 0.1)$, the hidden sector is
colder relative to the visible sector. As shown in
Ref.~\cite{Feng:2008mu}, a colder hidden sector requires a smaller
$\alpha_X$ to meet the relic abundance constraint, which results in a
smaller momentum transfer rate.  Furthermore, for a colder hidden
sector, the visible sector appears hotter. The resulting larger Hubble
expansion rate of \eqref{eq:hubble} also causes the momentum transfer
processes to become inefficient earlier.  This effect is even more
pronounced for $(\tantheta, \xiRH) = (10, 0.1)$, since large
$\tantheta$ implies even lower $\alpha_X$, increasing the visible
sector $T_{\text{kd}}$ further.

\section{Power Spectrum of Dark Matter Density Fluctuations}
\label{sec:matter}

The interactions and subsequent decoupling of dark matter particles
damp the matter power spectrum. The matter power spectrum is the
Fourier transform of the two-point function of the cosmological
density fluctuations in matter. For Gaussian initial conditions, this
encodes all the relevant information for linear perturbation
theory. In this section, we will study the power spectrum of hidden
charged dark matter to contrast its predictions with that of canonical
WIMP models. We will focus on small scales where differences might be
expected, especially in the predictions for the mass of the smallest
dark matter structures that can form. We will find that the
differences from the usual WIMP models are not observable with current
data, although these may be observable in the future.  However, the
smallest halo mass prediction is important for the calculation of
Sommerfeld-enhanced self-annihilations, which we will discuss in the
following section.

The damping due to the coupling of charged dark matter to hidden
sector photons (and neutrinos) results from two distinct
processes. The coupling of the dark matter to the hidden sector
photons and neutrinos leads to damped oscillatory
features~\cite{Loeb:2005pm} due to the interplay between the gravity
of dark matter and the pressure of the photons. This is the dominant
effect for the case of WIMPs with electroweak scale masses and also
for our case with $m_X > 1~\gev$. In addition, after decoupling, the
free-streaming of the dark matter particles further suppresses the
power spectrum. This effect dominates at lower masses and in
particular in the region of parameter space where interactions with
the hidden sector neutrinos dominate over those with hidden sector
photons.

The presence of the hidden sector and the coupling of the dark matter
to $\gamma^h$ and $\nu^h$ make the decoupling and matter power
spectrum calculation different from the standard WIMP case. The two
main differences are the following. In the standard WIMP case, the
WIMP couples to standard model fermions that are part of a tightly
coupled (collisionally-coupled) fluid. This implies that multipole
moments of the density fluctuations higher than the dipole are
strongly suppressed. This result stems from the fact that in a
collisionally-coupled plasma, a quadrupole anisotropy pattern can only
develop if there is significant diffusion. In the present case, the
hidden sector photons and neutrinos are only coupled to the dark
matter and are able to diffuse significantly.  We therefore need to
track their density fluctuations to higher multipoles. The other
notable difference is that the expansion rate is predominantly set by
the visible sector matter and so the ratio of the scattering rate in
the hidden sector to expansion rate is different from the usual WIMP
case. We provide details about this calculation in \appref{app}.

\begin{figure}[tb]
\begin{center}
\includegraphics*[scale=1.4]{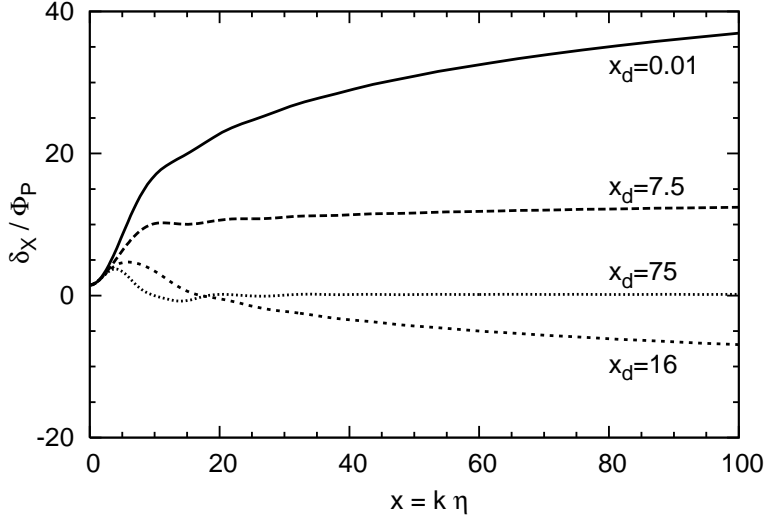}
\end{center}
\vspace*{-.29in}
\caption{The normalized amplitudes of dark matter fluctuation for
different modes with comoving wavenumbers $x_d=0.01,7.5,75,16$ as
functions of $x=k\eta$, where $\eta$ is the conformal time. We fix
$(\tantheta, \xiRH) = (\sqrt{3/5}, 0.8)$ for this plot.}
\label{fig:delta}
\end{figure}

In \figref{delta} we show the normalized density fluctuations for
different modes as functions of $x = k \eta$, where $\eta$ is the
conformal time. The coupling between $\tilde{\tau}^h$ and $\gamma^h$
results in damped oscillations as mentioned previously, and this is
apparent in \figref{delta} for the modes that enter the horizon before
kinetic decoupling. However, there is one major difference between
this hidden sector scenario and the WIMP scenario studied in
Refs.~\cite{Loeb:2005pm,Bertschinger:2006nq}: the oscillation
amplitude is much smaller than it is in the WIMP case. This traces
back to the $\tilde{\tau}^h \gamma^h$ scattering cross section, which
is $\sim \alpha_X^2/m_X^2 \sim g_W^4 / m_W^2$, that is, set by the SM
weak scale, which is far smaller than the analogous visible sector
cross section, which is the Thomson cross section $\sim g_W^4/m_e^2$.
The smaller opacity in the hidden sector case results in large
diffusion damping and hence the oscillations are highly suppressed.

\begin{figure}[!p]
\begin{center}
\includegraphics*[scale=1.4]{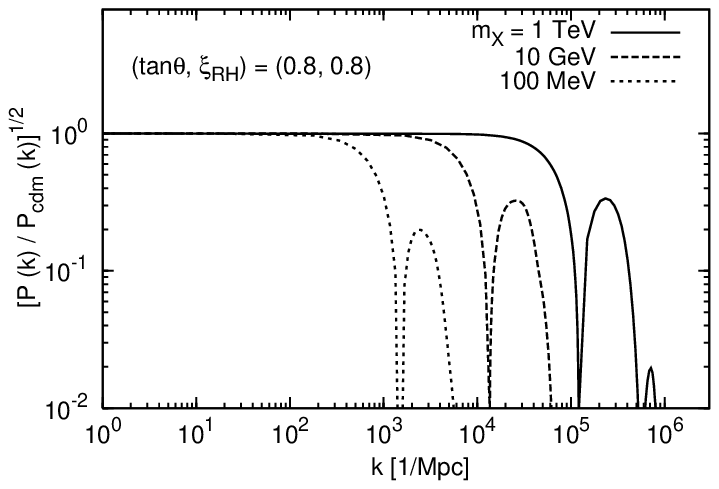} \\
\vspace*{-.15in}
\includegraphics*[scale=1.4]{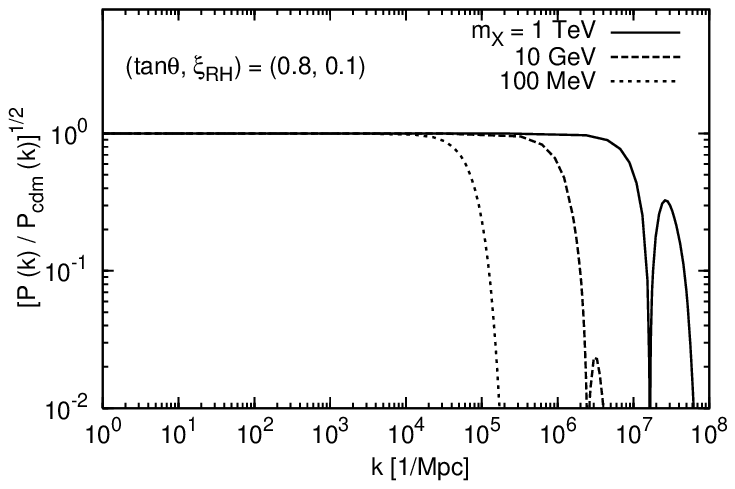} \\
\vspace*{-.15in}
\includegraphics*[scale=1.4]{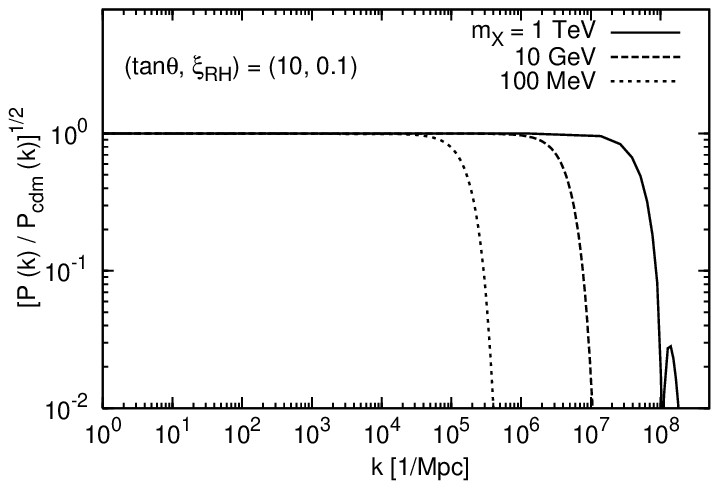}
\end{center}
\vspace*{-.29in}
\caption{Transfer functions of the normalized dark matter density
perturbation amplitude for $(\tantheta, \xiRH) = (\sqrt{3/5},
0.8)$ (top), $(\sqrt{3/5},0.1)$ (middle), and $(10, 0.1)$ (bottom).  }
\label{fig:MPfor0.8}
\end{figure}

In \figref{MPfor0.8} we plot transfer functions of the density
perturbation of the hidden charged dark matter for various $m_X$ and
combinations of $(\tantheta, \xiRH)$.  For smaller masses, kinetic
decoupling happens later and the matter power spectrum is more
suppressed. We define the cut-off wavenumber $\kcut$ as the point
where the transfer function first drops to $1/e$ of its value at small
wavenumbers.  For $(\tantheta, \xiRH) = (\sqrt{3/5}, 0.8)$ (top panel)
and $m_X=1~\tev$, 10 GeV and 100 MeV, we find $\kcut = 8.0 \times
10^4~\Mpc^{-1}$, $9100~\Mpc^{-1}$, and $970~\Mpc^{-1}$, respectively.
The free-streaming damping scale is
\begin{eqnarray}
\lambda_{\rm fs}\approx\left[\Gamma \left(\frac{1}{2} \right)
\frac{T^h_{\rm kd}}{m_X}\right]^{\frac{1}{2}}
\eta_d\ln\left(\frac{\eta_{\rm eq}}{\eta_d}\right) \ ,
\end{eqnarray}
where $\eta_d$ is the comoving horizon at the time of kinetic
decoupling, and $\eta_{\rm eq}$ is the comoving horizon at
matter-radiation equality, and $\Gamma\left(1/2\right)=\sqrt{\pi}$.
For $m_X=1~\tev$, 10 GeV, and 100 MeV, we find $\lambda^{-1}_{\rm
fs} = 9.3 \times 10^5~\Mpc^{-1}$, $3.8 \times 10^4~\Mpc^{-1}$, and
$1500~\Mpc^{-1}$, respectively. We see that for these masses,
$\lambda_{\rm fs} < k_{\rm cut}^{-1}$ --- the cut-off scale is
determined by the acoustic damping.

In \figref{MPfor0.8}, we also show the transfer function of the dark
matter density perturbation for the cases $(\tantheta, \xiRH) =
(\sqrt{3/5},0.1)$ (middle panel), and $(10, 0.1)$ (bottom panel).
Comparing with the top panel, we can see the transfer function cuts
off at larger wavenumber for colder hidden sectors. Kinetic decoupling
happens earlier and only the small scale modes that entered the
horizon before kinetic decoupling get suppressed. For $m_X = 1~\tev$,
the cut-off is determined by the acoustic damping scale. However, for
$m_X = 10~\gev$ and 100 MeV, the free-streaming damping scale is
comparable to the acoustic damping scale.

The matter power spectrum contains all the information about linear
Gaussian density fluctuations. These fluctuations are amplified by
gravity to create non-linear structures, that is, dark matter
halos. However, on scales below the cut-off the matter distribution is
smooth and gravity cannot regenerate power on these scales. Thus the
linear power spectrum also encodes information about the smallest
building blocks of dark matter halos. The minimal mass of dark matter
clumps may be estimated as
\begin{equation}
\Mcut = \frac{4\pi}{3}\left(\frac{\pi}{\kcut}\right)^{3}
\Omega_m \rho_{\rm crit} \ ,
\end{equation}
where $\Omega_m$ is the matter density today, and $\rho_{\rm crit} = 3
H^2_0 / (8 \pi G) = 8.1 h^2 \times 10^{-47}~\gev^4$ is the present
critical density.

\begin{figure}[tb]
\begin{center}
\includegraphics*[scale=1.4]{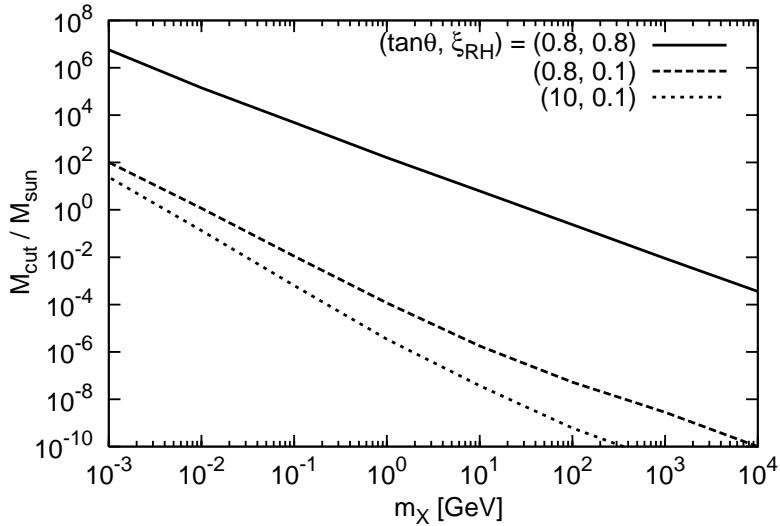}
\end{center}
\vspace*{-.29in}
\caption{Mass of the smallest virialized dark matter structure that
  can form, as a function of the dark matter mass $m_X$ for
  $(\tantheta, \xiRH) = (\sqrt{3/5}, 0.8), (\sqrt{3/5},0.1), (10,
  0.1)$, as indicated.}
\label{fig:Mc}
\end{figure}

In \figref{Mc} we show the characteristic mass of the smallest objects
as a function of dark matter mass $m_X$ for various values of
$(\tantheta, \xiRH)$. For the lower reheating temperature parameter
$\xiRH$, $\Mcut$ is smaller by a few orders of magnitude for a given
dark matter mass. For cold hidden sectors, kinetic decoupling occurs
earlier, corresponding to smaller comoving horizons and larger
$\kcut$, which leads to smaller $\Mcut$.  We can understand this more
quantitatively. The parameter $\kcut$ is related to the comoving
horizon when kinetic decoupling happens as $\kcut \eta_d= x_d$, where
a typical value is $x_d \sim 5.9$ in our case. On the other hand, we
have $\eta_d \sim 1/T_{\text{kd}}$, so $\kcut$ is proportional to the
decoupling temperature, that is, $\kcut \sim T_{\text{kd}}$, and
therefore $\Mcut \sim (T_{\text{kd}})^{-3}$. As an example,
consider the $m_X=1~\tev$ case. {}From \figref{Tkd}, the ratio of
visible decoupling temperatures for the $(\tantheta, \xiRH) =
(\sqrt{3/5}, 0.8)$ and $(\sqrt{3/5}, 0.1)$ cases is $7\times 10^{-3}$,
and so the expected ratio of minimum halo masses is $3\times 10^6$ as
may be ascertained from \figref{Mc}.

Our results in \figref{Mc} show that the minimum halo mass --- the
building blocks of structure in the universe --- could range anywhere
from $10^{-10} M_{\odot}$ to the size of the smallest galaxies
observed in the universe for dark matter masses within the MeV to TeV
range.  We will see in \secref{halo} that if we restrict our attention
to $\tantheta < 10$, we will require $m_X \agt 1~\gev$ to obtain the
right relic density and be consistent with the observed morphology of
galactic dark matter halos. For this range of parameter space, we see
from \figref{Mc} that the minimum mass halos are less than about $10^4
M_\odot$. At the present time, there is no way to test for halo masses
as small as this.  These predictions for structure formation are
therefore indistinguishable from those of WIMP models, and current
observations do not place constraints on the hidden charged dark
matter parameter space.

In the future, it may become possible to probe the upper end of this
range with strong gravitational lensing of radio-loud
quasars~\cite{Hisano:2006cj}. In addition, if the dark sector is
linked to the visible sector through connector particles with both
visible and hidden sector quantum
numbers~\cite{Feng:2008ya,Feng:2008dz,Feng:2008qn}, then
self-annihilations of dark matter within these mini-halos might
provide a detectable signature for the Fermi gamma-ray
experiment~\cite{Diemand:2005vz,Ando:2005xg,Koushiappas:2006qq,%
Ando:2006cr,Cuoco:2007sh,Ando:2008br,SiegalGaskins:2008ge,Lee:2008fm}.

\section{Bound State Formation and Sommerfeld-Enhanced Annihilations
in the Early Universe}
\label{sec:boundstate}

Charged dark matter annihilation may be enhanced in two ways.  First,
at low velocities, its annihilation cross section is enhanced by the
Sommerfeld effect~\cite{Sommerfeld:1931}; in Feynman diagrammatic
language, this enhancement arises from diagrams with additional
$\uone$ gauge bosons exchanged in the $t$-channel that are higher
order in gauge coupling, but kinematically enhanced at low
velocities~\cite{Hisano:2002fk,Hisano:2003ec,Hisano:2004ds}. Second,
charged dark matter may form bound states, which then annihilate.
These are independent effects that must be included separately.  Both
act to enhance annihilation, and one might worry that they have
negative implications.  For example, the resulting enhanced
annihilation rates may reduce the thermal relic density after chemical
freeze out to negligible levels or otherwise change the relic density
in a way that is excluded by astrophysical observations.

In full generality, these effects may be evaluated by solving the
coupled Boltzmann equations for the number densities of the bound
state $B \equiv (\tilde{\tau}^{h+} \tilde{\tau}^{h-})$ and the dark
matter particle $X \equiv \tilde{\tau}^h$,
\begin{eqnarray}
\label{Boltzmann}
   \frac{d n_B}{dt} + 3 H n_B &=&
   n_X \Gamma_{\rm rec}
   - n_B \Gamma_B - n_B \Gamma_{\rm ion} \nonumber \\
   \frac{d n_X}{dt} + 3 H n_X &=&
   - n_X \Gamma_{\rm ann} - n_X \Gamma_{\rm rec}
   + n_B \Gamma_{\rm ion} \ ,
\end{eqnarray}
where
\begin{eqnarray}
\Gamma_{\text{rec}} &\equiv& \Gamma ( X^+ X^- \to B \gamma^h)
= C\, n_X\, \langle \sigma_{\text{rec}} v \rangle \nonumber \\
\Gamma_B &\equiv& \Gamma (B \to \gamma^h \gamma^h )
= \frac{\alpha^5_X\, m_X}{2} \nonumber \\
\Gamma_{\text{ion}} &\equiv& \Gamma ( B \gamma^h \to X^+ X^- )
= n_{\gamma^h}\, \langle \sigma_{\text{ion}} v \rangle \nonumber \\
\Gamma_{\text{ann}} &\equiv& \Gamma ( X^+ X^- \to \gamma^h \gamma^h )
= C\, n_X\, S\, \langle \sigma_A v \rangle \ ,
\label{rates}
\end{eqnarray}
are the recombination (bound state formation) rate, the bound state
decay width, the ionization rate, and the Sommerfeld-enhanced dark
matter annihilation rate, respectively.  In \eqref{rates}, $C=\langle
\rho^2\rangle/\langle \rho \rangle^2$ accounts for the clumping of
dark matter into halos, $\sigma_A$ is the tree-level $S$-wave
annihilation cross section, and
\begin{equation}
S = \frac{\pi \alpha_X / v}{1 - e^{-\pi \alpha_X /v}}
\label{Sommerfeldfactor}
\end{equation}
is the Sommerfeld enhancement
factor~\cite{Sommerfeld:1931,Cirelli:2007xd}, where $v$ is the
center-of-mass velocity of the incoming particles.  For $v \agt
\alpha_X$, $S \sim 1$, but for $v \ll \alpha_X$, $S \sim \pi \alpha_X
/ v$ may be a significant enhancement.  The recombination and
photoionization and cross sections are related through the Milne
relation
\begin{equation}
   \frac{\sigma_{\rm ion}}{\sigma_{\rm rec}} = \left(\frac{m_X\, v}
   {B_n + \frac{1}{2} m_X\, v^2} \right)^2\,
   \frac{g^2_{\tilde{\tau}}}{2\,g_n}\ ,
\end{equation}
where $g_{\tilde{\tau}}=1$ and $g_n=1$ are the statistical weights of
the staus and the $n$-th level of the bound state, and $B_n =
\alpha^2_X\, m_X / 4 n^2$ is the binding energy of bound state level
$n$.  Photoionization may, in principle destroy bound states before
they annihilate.

The recombination rate requires some analysis.  First, note that a
photon is radiated in the recombination process; bound state formation
followed by bound state decay yields a different final state than
Sommerfeld-enhanced annihilation, and these two effects must be
included separately; as we will see, both yield an enhancement that is
parametrically $ \sim \alpha_X / v$ for small velocities.

The recombination cross section may be determined by using a hydrogen
atom wavefunction solution in the Schr\"odinger equation for a
particle moving in a central potential $V (r) \sim - \alpha_X / r$.
The recombination cross section for the $n$-th shell
is~\cite{Bethe:1957}
\begin{equation}
\label{xsec_recombination}
   \sigma^{(n)}_{\rm rec} = \frac{2^8 \pi^2}{3}\, \frac{\alpha_X}{m_X}\,
   \frac{1}{m_X}\, \frac{B_n}{m_X v^2}\,
   \left(\frac{B_n}{B_n + \frac{1}{2} m_X v^2} \right)^2\,
   \cdot\, \frac{e^{-4 \sqrt{\frac{2 B_n}{m_X v^2}}\,
   \tan^{-1} \sqrt{\frac{m_X v^2}{2 B_n}}}}{1 - e^{-2 \pi\,
   \sqrt{\frac{2 B_n}{m_X v^2}}}}\, .
\end{equation}
This result includes a correction factor from the fact that the
electromagnetic force is long range, so that the incoming state is not
a plane wave.  The phase space distribution of the $\tilde{\tau}^h$
cold dark matter is given by solving the Boltzmann equation of
\eqref{eq:kinetic_coupling}.  The solution is a Boltzmann distribution
with dark matter temperature $T_X \sim T^h$ before kinetic decoupling,
and an effective temperature
\begin{equation}
\label{eq:Tcdm_asymptotic}
   T_X \to \frac{T^{h\, 2}}{m_X}\, \frac{m_X}{T^h_{\text{kd}}}
\end{equation}
after kinetic decoupling.  The $\tilde{\tau}^h$ thus has Maxwellian
velocity distribution at the effective temperature $T_X$,
\begin{equation}
   f (v)\, dv = 4 \pi \left(\frac{m_X}{2 \pi T_X} \right)^{3/2}\, v^2\,
   e^{- \frac{m_X v^2}{2 T_X}}\, dv \ .
\end{equation}
Summing over all binding states, the thermally averaged total
recombination cross section is
\begin{eqnarray}
\label{rec_thermalaverage}
   \langle \sigma_{\rm rec} v \rangle
   &=& \sum_n \int dv f(v)\, \sigma^{(n)}_{\rm rec}\, v \nonumber \\
   &=& \sum_n \frac{4}{\sqrt{\pi}}\,
   \frac{2^8 \pi^2}{3}\, \frac{1}{v^3_0}\, \frac{\alpha_X}{m^3_X}\,
   B_n\, \int dv\, v\, \frac{B^2_n\, e^{-v^2/v^2_0}}{(B_n +
   \frac{1}{2} m_X v^2)^2}\,
   \frac{e^{-4 \sqrt{\frac{2 B_n}{m_X v^2}}\,
   \tan^{-1} \sqrt{\frac{m_X v^2}{2 B_n}}}}{1 - e^{-2 \pi\,
   \sqrt{\frac{2 B_n}{m_X v^2}}}}\, \nonumber \\
    &\to& \frac{(4 \pi\, \alpha_X)^2}{m^2_X}\, \frac{1}{8 \pi}\,
    \frac{\pi \alpha_X}{v_0}\,
    \left(\frac{2^6\, e^{-4}}{3 \sqrt{\pi}} \right)\,
    \left(\sum_n \frac{1}{n^2}  \right)\, ,
    \qquad \frac{v_0}{\alpha_X} \ll \frac{1}{n}\, ,
\end{eqnarray}
where we have made the replacement $v_0^2 = 2\, T_X / m_X$.  We see
that in the small kinetic energy limit, the thermally averaged
recombination cross section is the thermally averaged annihilation
cross section of \eqsref{partial_wave}{partial_wave2}, enhanced by a
``bound state enhancement factor,'' that is, $\langle
\sigma_{\text{rec}} v \rangle = S_{\text{rec}}\, \langle \sigma_A v
\rangle$, with
\begin{equation}
S_{\rm rec} \sim 0.7\, \left(\frac{1}{8 \pi\, a_0} \right)\,
\left(\frac{\alpha_X}{v_0} \right)\, , \,
\qquad v_0 / \alpha_X \ll 1 \ .
\end{equation}
The numerical prefactor is, for example, 0.5 and 0.05 for $\tan
\theta^h_W = \sqrt{3/5}$ and 10, respectively.  Summing to higher
binding states would give a factor of $\simeq 1.6$ at most.

\begin{figure}[tb]
\begin{center}
\includegraphics*[scale=1.4]{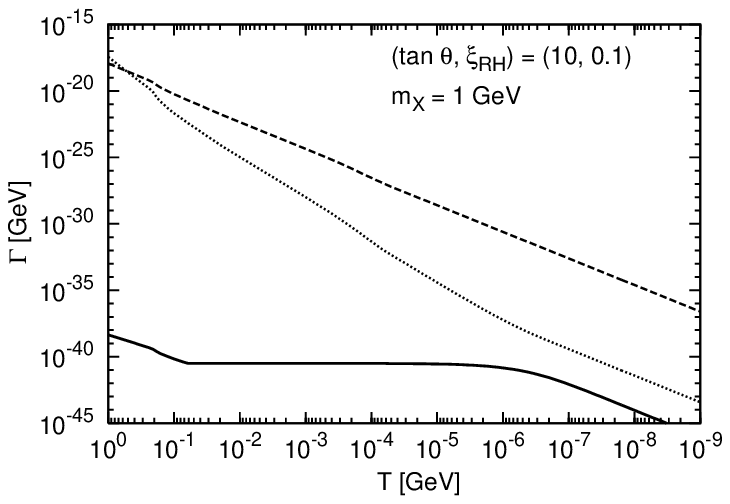} \\
\vspace*{-.15in}
\includegraphics*[scale=1.4]{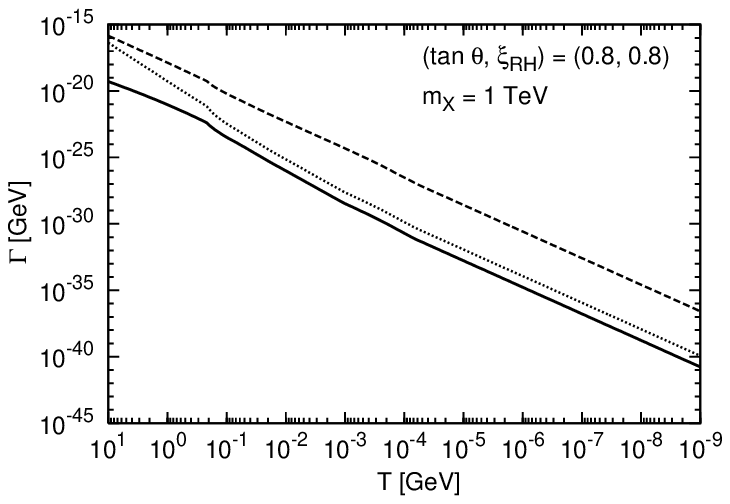}
\end{center}
\vspace*{-.29in}
\caption{The recombination rates (bound state formation rates)
$\Gamma_{\rm rec}$ (solid) in the early universe as functions of the
visible (CMB) temperature for dark matter mass $m_X = 1~\gev$ and
$(\tantheta, \xiRH) = (10, 0.1)$ (upper) and $m_X = 1~\tev$ and
$(\tantheta, \xiRH) = (\sqrt{3/5}, 0.8)$ (lower). Also shown are the
Sommerfeld-enhanced annihilation rates $\Gamma_{\rm ann}$ (dotted) and
the Hubble expansion rate $H$ (dashed).}
\label{fig:boundrate}
\end{figure}

In \figref{boundrate} we plot the bound state formation rates
$\Gamma_{\rm rec}$ for $m_X = 1~\gev$ and 1 TeV.  After chemical
freeze out and before the bound state formation can take place, the
cold dark matter number density is $n_X = \Omega_X\, \rho_{\rm crit}\,
a^{-3} / m_X \propto T^3$. Here we compare in the era before
matter-radiation equality where the dark matter particles have not
clumped into halos yet.  Next we will consider enhanced annihilation
and bound state formation in the earliest bound objects, the
protohalos.  The bound state formation rate for different dark matter
masses $m_X$ may be understood as follows.  In general, there are
three stages:
\begin{enumerate}
\item Before kinetic decoupling the dark matter most probable speed is
$v_0 \propto T^{1/2}$. When $v_0 /\alpha_X \gg 1$ the thermally
averaged bound state formation cross section goes as $\langle
\sigma_{\rm rec} v \rangle \propto 1/v^3_0$ and hence the rate
$\Gamma_{\rm rec} \propto T^{3/2}$.
\item After kinetic decoupling, the dark matter most probable speed
becomes $v_0 \propto T$. When $v_0 /\alpha_X \gg 1$ still holds, the
bound state formation rate is $\Gamma_{\rm rec} \propto T^3\, v^{-3}_0
= {\rm constant}$.
\item When the dark matter velocity drops to the point $v_0 /\alpha_X
\alt 1$, $\langle \sigma_{\rm rec} v \rangle$ becomes $\propto 1/v_0$
(cf. \eqref{rec_thermalaverage}).  As a consequence $\Gamma_{\rm rec}$
decreases as $T^2$, the same as the Hubble rate in the
radiation-domination era.
\end{enumerate}

Also shown in \figref{boundrate} are the Hubble expansion rate $H$ and
the Sommerfeld-enhanced annihilation rates $\Gamma_{\rm ann}$.  The
results can be summarized as
\begin{equation}
   H \gg \Gamma_{\rm ann} \agt \Gamma_{\rm rec}
\end{equation}
for $m_X = 1~\gev - 10~\tev$.  The bound state decay rate $\Gamma_B$
and photoionization rate $\Gamma_{\rm ion}$ can be much larger than
the Hubble rate.  But since bound states rarely form, we may set $n_B
= 0$ in \eqref{Boltzmann}.  We see that for $v \ll \alpha_X$, bound
state formation has qualitatively the same effect on annihilation as
the Sommerfeld effect --- they enhance annihilation by factors of $S$
and $S_{\text{rec}}$, which are both proportional to $\alpha_X/v$,
with order one coefficients.  Since the Sommerfeld enhancement factor
is also valid for $v \agt \alpha_X$, while the bound state effect
becomes negligible at large $v$, one may draw qualitatively correct
conclusions by considering only the Sommerfeld effect and neglecting
the effects of bound state formation, but we emphasize that these
effects are independent, yielding two different final states.

We may now answer the questions posed at the beginning of this
section.  Because the Hubble expansion rate is larger than the
Sommerfeld-enhanced and bound state-catalyzed annihilation rates, the
thermal relic density is not modified significantly between the times
of chemical freeze out and matter-radiation equality.

\section{Sommerfeld-Enhanced Annihilations in Protohalos}
\label{sec:protohalo}

We now consider these non-perturbative effects on the dark matter
annihilation after matter-radiation equality, when the growth of
structure enters the non-linear regime.  The smallest structures
undergo gravitational collapse first and the more massive ones form
later.  The minimum halo mass $\Mcut$ is set by the cold dark matter
kinetic decoupling temperature and we have computed this for the
hidden charged dark matter in \secref{matter} (cf. \figref{Mc}).  The
redshift at which these objects form (virialize) depends on the
underlying matter power spectrum~\cite{Navarro:1996gj}, which we have
also computed in \secref{matter}.  Simulations~\cite{Diemand:2006ey}
find $z_c \sim 30$ for a typical $10^{-6}\, M_\odot$ halo and we will
assume this value in this work, but note that in detail the collapse
redshift depends on $\Mcut$.  We take the overdensity of the
virialized region to be $\simeq 178$ times the ambient cosmological
density at $z_c$ (appropriate for WMAP cosmology) and estimate the
velocity dispersion of the dark matter particles in the halo as:
\begin{equation}
   v_0 = \sqrt{\frac{G M_{\rm cut}}{R_{\rm vir}}} \simeq
   3.0 \times 10^{-8} \left(\frac{\Mcut}{M_\odot} \right)^{1/3}\,
   \left(1 + z_c \right)^{1/2}\, .
\end{equation}
Hidden charged dark matter particles in our model with masses $m_X
\agt 1~\gev$ have velocity dispersion $v_0 \ll \alpha_X$ in these
protohalos.  They thus form bound states and pair-annihilate with an
enhanced cross section, from an initial number density $n_X \simeq
6~(m_X /\gev)^{-1}~\cm^{-3}$ for about a Hubble time before the
protohalos merge into larger halos.  In these larger halos, the dark
matter particles have smaller phase space density and so the
annihilation rate is lower.

Dark matter annihilation in protohalos was used in
Ref.~\cite{Kamionkowski:2008gj} to obtain stringent constraints on
models with the Sommerfeld enhancement effect.  In that work dark
matter particles were assumed to annihilate dominantly to visible
photons and $e^+ e^-$ pairs, which contribute to the extragalactic
diffuse gamma-ray flux and generate CMB temperature and polarization
power spectra. However, in the present context, where the hidden
sector is not coupled to the visible sector, only a much weaker,
separate, constraint from CMB anisotropy measurements applies.  The
CMB measurement of the matter density $\Omega_m h^2$ can be compared
with those obtained by distance measurements using Supernovae Type Ia
(SNIa) and baryon acoustic oscillations (see, for example,
Ref.~\cite{Komatsu:2008hk}). A rough estimate will suffice for our
purposes here and we will assume that no more than 10\% of the total
matter can be converted into radiation between the epoch of
recombination and now.

This constraint may then be applied to our models, where the $S$-wave
stau annihilation cross section of \eqref{partial_wave} is enhanced by
the Sommerfeld factor $S$.  As mentioned before, when $v_0 /\alpha_X
\ll 1$ the bound state formation cross section is of comparable size.
One demands the fraction of stau annihilation in the protohalos
\begin{equation}
    f \simeq n_X\, S\, \langle \sigma_A\, v \rangle\, C\, t
    \simeq \frac{178\, \Omega_c\, \rho_{\rm crit}\, \left(1+z_c \right)^3}
    {m_X}\, S\, \langle \sigma_A\, v \rangle\, C\, t\, \leq\, 0.1\, ,
\label{fbound}
\end{equation}
during $t \simeq 5.6 \times 10^{17}\, (1+z_c)^{-1.5}~\s$, the age of
the universe at redshift $z_c \gg 1$.  The density profile of the
minimum mass halos~\cite{Diemand:2006ey} is expected to be similar to
those of the larger halos~\cite{Navarro:1996gj}, and we include an
appropriate clumping factor $C = 7$ for the minimum halo mass.  We
find that for $(\tantheta, \xiRH) = (10, 0.1)$, $f \sim 10^{-8} -
10^{-5}$ for $m_X=1~\gev - 10~\tev$ and $\alpha_X$ fixed by $\Omega_X
h^2 = 0.11$.  For $(\tantheta, \xiRH) = (\sqrt{3/5}, 0.8)$ and
$(\sqrt{3/5}, 0.1)$, $f \sim 10^{-7} - 10^{-5}$ for the same mass
range and the corresponding $\alpha_X$'s.  These values are far below
the bound of $f \leq 0.1$, and so the annihilation of dark matter in
protohalos does not constrain the scenario in the case of purely
hidden charged dark matter.  However, if there were connector
particles with both hidden and visible sector quantum numbers
mediating annihilation to observable particles, a much more stringent
bound $f \alt 10^{-9}$ may apply~\cite{Kamionkowski:2008gj}.  This
is from the consideration that (visible) photons ejected at an energy
$\agt 300~\gev$ or $\alt 100~\kev$ should not lead to heating
and ionization of the intergalactic medium that contradicts CMB and
large scale structure observational data.

More massive halos have larger dark matter velocity dispersions. For
example, Milky Way-size halos have $v_0 \sim 270~{\rm km/s} \sim
10^{-3}$, while dwarf galaxies can have velocities as low as $v_0 \sim
10~{\rm km/s} \sim 7 \times 10^{-5}$.  For hidden charged dark matter
particles with masses $m_X = 1~\gev - 10~\tev$, then, only a fraction
$f \sim 10^{-7}$ annihilates during the age of the Universe.

\section{Ellipticity and Cores of Dark Matter Halos}
\label{sec:halo}

At late times, elastic scattering between $\tilde{\tau}^h$ dark matter
particles in the halos of galaxies or clusters of galaxies may change
the shape of constant-density contours and lead to the formation of a
core, that is, a central region with constant density. The main effect
results from transfer of kinetic energy in collisions and if this
process is fast enough to create ${\cal O}(1)$ changes to the energy
of the dark matter particles in the halo, then it will drive the halo
towards isothermality and isotropize the mass distribution.

The true dynamical picture is more complicated. Initially, heat is
conducted from the hotter outer to the cooler inner parts of the halo
through collisions. This heats up the core causing it to puff up,
which gives rise to a density profile that is much shallower (or even
flat) compared to the initial cuspy density profile of the central
regions of the halo. In addition, self-annihilations will also lead to
cores~\cite{Kaplinghat:2000vt}, however our calculation in
\secref{protohalo} has shown that this is a not large effect.
The collisions also erase velocity correlations
and lead to more a more spherical (rounder) core. We find that the
observed ellipticity of galactic dark matter halos provides the
strongest constraints on charged dark matter models.

Over periods long compared to the relaxation time, ejection of dark
matter particles from the parent halo in collisions will further cool
the core. For an isolated halo, this will eventually lead to core
collapse and result in a steep cusp.  In a cosmological setting,
however, halos accrete dark matter particles and this would offset the
energy loss due to collisions. Cosmological
simulations~\cite{Dave:2000ar,Yoshida:2000bx} seem to indicate that
core collapse occurs on a time scale much longer than the relaxation
time scale, if at all. We will, therefore, not consider this effect
any further here.

Shapes of dark matter halos of elliptical
galaxies~\cite{Buote:2002wd,Humphrey:2009} and
clusters~\cite{MiraldaEscude:2000qt,Fang:2008ad} are decidedly
elliptical and this fact may be used to put constraints on the
self-interactions. Such an analysis~\cite{MiraldaEscude:2000qt} was
carried out in the context of the self-interacting dark matter
proposal~\cite{Spergel:1999mh}. This proposal was designed to explain
why observations of the rotation curve of low surface brightness
galaxies seemed to indicate that the dark matter distribution had a
flat density profile in the inner core~\cite{Moore:1994yx,de %
Blok:1997ut,Moore:1999gc} and to explain the observed census of dwarf
galaxies in the Local Group~\cite{Klypin:1999uc,Moore:1999nt}. The
status of these discrepancies is unclear --- while it seems clear that
the census of the observed local group galaxies are broadly in
agreement with the cold dark matter
model~\cite{Strigari:2007ma,Tollerud:2008ze}, the fit to rotation
curves of low-surface brightness galaxies is still
problematic~\cite{Simon:2004sr,Gentile:2004tb,Gentile:2005tu,%
Spano:2007nt,KuziodeNaray:2007qi}.

The strongly self-interacting dark matter proposal motivated the first
numerical simulations to deduce the effects of dark matter
self-interaction~\cite{Hannestad:1999pi,Kochanek:2000pi,Yoshida:2000uw,%
Dave:2000ar,Wandelt:2000ad,Moore:2000fp,Craig:2001xw,Ahn:2004xt,%
Randall:2007ph} and they validated the expectation that during the
evolution to isothermality, the dark matter core becomes rounder.
These simulations indicated that only regions of parameters space with
$500~\gev^{-3} \alt \sigma_{\rm DM} / m_X \alt 5000~\gev^{-3}$
($0.1~\cm^2/\g \alt \sigma_{\rm DM} / m_X \alt 1~\cm^2/\g$) could
introduce observable features on the scale of the dwarf galaxies while
at the same time being consistent with observations of larger
galaxies. In the constraint quoted above, $\sigma_{\rm DM}$ is the
dark matter elastic self-scattering cross-section modeled as a
hard-sphere interaction. The analysis of the shape of the dark matter
halo of a particular cluster of galaxies (using gravitational lensing)
indicated that much of the above preferred region was ruled
out~\cite{MiraldaEscude:2000qt}.

We revisit the constraint from the inferred ellipticity of dark matter
halos in the rest of this section. These constraints arise from a
wide-range of observations including
X-rays~\cite{Buote:2002wd,Fang:2008ad}, strong
lensing~\cite{Inada:2003vc,Koopmans:2006iu,Morokuma:2006qq} and weak
lensing~\cite{Hoekstra:2003pn,Mandelbaum:2005nf,Evans:2008mp}. Recent
work on combining different measurements to reveal the anisotropy of
velocity dispersion~\cite{Host:2008gi} in clusters (as opposed to the
shape of dark matter distribution) is a different but
complementary way to constrain the self-interaction of dark
matter.

To estimate how these observations may be used to constrain the dark
sector Coulomb interactions, we calculate the relaxation time for
establishing an isothermal halo. We will then assume that the time
scale for isotropizing the spatial distribution of the dark matter
halo is the same as this relaxation time and use constraints from
measurements of the  ellipticity of galaxy halos to put limits on the
dark sector Coulomb interaction. The notable feature of the Coulomb
interaction is the strong dependence of the energy transfer rate on
the relative velocity of the interacting particles. This ties in with
other investigations that considered velocity-dependent interaction
cross sections. See, for example, Refs.~\cite{Firmani:2000qe,%
Firmani:2000ce,Hennawi:2001be,Colin:2003ed,SanchezSalcedo:2005vc,Koda}.

The Rutherford scattering cross section is
\begin{equation}
   \frac{d \sigma}{d \Omega} = \frac{\alpha^2_X}{4 m^2_X v^4
   \sin^4 \left(\theta/2 \right)} \ ,
\end{equation}
where $\theta$ is the scattering angle in the lab frame. We assume
the dark matter particles in the halos have a local Maxwellian
velocity distribution $f(v)$ (normalized to unity) and velocity
dispersion $\langle v^2 \rangle =(3/2) v_0^2(r)$, which in general
varies spatially within a halo.  The kinetic energy exchange in each
collision is $\delta E_k = E_k \left(1 - \cos \theta \right)$, where
$E_k =m_X v^2 / 2$.  The rate of energy transfer is, then,
\begin{eqnarray}
   \dot{E_k} &=& \int d v\, d \Omega \frac{d \sigma}{d \Omega}
   f(v)\, \delta E_k\, v\, n_X\, \nonumber \\
   &=& 2 \pi\, \frac{\alpha^2_X}{4 m^2_X}\, \frac{4 }{\sqrt{\pi}}\,
   \frac{1}{v^3_0}\, \frac{1}{2} m_X\, \frac{\rho_X}{m_X}\,
   \int dv\, d \cos\theta \frac{1-\cos\theta}{\sin^4
   \left(\theta/2 \right)}\, v\, e^{-v^2/v^2_0} \nonumber \\
   &=& - \frac{2 \alpha^2_X \rho_X\, \sqrt{\pi}}{m^2_X\, v^3_0}\,
   \ln (1 - \cos\theta_{\rm min})\, v_0^2\, .
\end{eqnarray}
The minimum scattering angle is related to the maximum impact
parameter through
\begin{equation}
   b_{\rm max} = \frac{\alpha_X}{m_X v^2_0}\, \cot
   \left(\theta_{\rm min} / 2 \right)\, ,
\end{equation}
where $b_{\rm max}$ should be chosen to be
\begin{equation}
   \lambda_D \sim \frac{m_X\, v_0}{\sqrt{4 \pi\, \alpha_X\,
   \rho_X}}\ ,
\end{equation}
the Debye screening length in the $\tilde{\tau}^h$ plasma.  We find
constraints on $\alpha_X$ by demanding that the relaxation time be
larger than the age of the universe,
\begin{equation}
\label{constraint_galdynamic}
  \tau_{\rm r} \equiv E_k / \dot{E}_k \simeq
  \frac{m^3_X\, v^3_0}{4 \sqrt{\pi}\,
  \alpha^2_X\, \rho_X}
  \left(\ln \left(
\frac{\left(b_{\rm max}\, m_\chi v^2_0 \alpha_X^{-1} \right)^2
+ 1}{2} \right)\right)^{-1}
  \geq 10^{10}\, \, {\rm years}\, ,
\end{equation}
where the ``Coulomb logarithm'' is $\sim 90$. The constraints on
$\alpha_X$ obtained above scale inversely with the square root of the
``phase space density'' $Q \equiv \rho_X /  v_0^3$, and this indicates
that the best constraints may be obtained by studying the cores of
galaxies rather than clusters of galaxies.

Many elliptical galaxies show clear evidence for flattened, triaxial
dark matter halos~\cite{Buote:2002wd}.  Using the profiles of the
total mass enclosed $M(r)$ and the halo concentration parameters $c$
from Refs.~\cite{Humphrey:2006rv,Humphrey}, we derive the radial
velocity dispersion $\overline{v^2_r} (r) = (3/2) v^2_0 (r)$ and the
dark matter density $\rho_X (r)$ at a radius $\sim 3 - 10$ kpc.  The
dark matter density drops from 3.5 to 0.7 GeV/cm$^3$ as one moves
outwards, while the velocity dispersion decreases slightly from 270 to
250 km/s.  Using these values in \eqref{constraint_galdynamic}, we
obtain a very stringent bound on $\alpha_X$. This constraint is shown
as the lower solid line in \figref{alpha_constraint}. If we demand
that the WIMPless scenario provide the right relic abundance of
$\Omega_X h^2 \simeq 0.11$, then for $(\tantheta, \xiRH) =
(\sqrt{3/5}, 0.8)$ and $(\sqrt{3/5}, 0.1)$, $m_X \agt 100~\gev$.
However, for $(\tantheta, \xiRH) = (10, 0.1)$, hidden charged dark
matter as light as $m_X \sim 1~\gev$ may freeze out with the correct
thermal relic density without being in conflict with the constraint
from elliptical halos.

The constraints obtained above are comparable to those deduced in
Ref.~\cite{Ackerman:2008gi}. This previous work demanded that the
properties of the Milky Way's dark matter halo (and by extension that
of other galaxies as massive as the Milky Way) should not deviate by
order unity from those predicted by collisionless dark matter
simulations. If we compare their result to our
\eqref{constraint_galdynamic} with $\rho_X=0.3~\gev~\cm^{-3}$ and
$v_0=200~\km~\s^{-1}$, we obtain weaker constraints on $\alpha_X$ by a
factor of $\sim 6$ at the same $m_X$.  We trace the bulk of this
discrepancy to the fact that they assume total dark matter mass of
$10^{10} M_\odot$ in their calculation and neglect the contribution of
stars in setting the local dark matter velocity dispersion.  The
constraints we have obtained from detailed observations of an
elliptical galaxy with strong X-ray emission resolved down to 3 kpc of
the center --- better than we can do within our own galaxy for now ---
are comparable to the bounds quoted in
Ref.~\cite{Ackerman:2008gi}. Future observations of hyper-velocity
stars within the Milky Way could strengthen these
constraints~\cite{Gnedin:2005pt}.

We may also consider dark matter halos more massive and less massive
than that of the Milky Way. Cluster halo shapes are measured with
X-rays and strong gravitational lensing at radii $\sim 100$ kpc.
We consider some Abell clusters for which the ellipticity profiles are
determined in Ref.~\cite{Fang:2008ad}.  The radial velocity dispersion
at this scale is typically $\sqrt{\langle v^2_r \rangle} \sim
1000~\km/\s$, as inferred from the $M(r)$ profiles determined in
Ref.~\cite{Vikhlinin:2005mp}.  As a consequence the bounds on
$\alpha_X$ are about two orders of magnitude weaker.

In principle, the most stringent constraint using
\eqref{constraint_galdynamic} would be  from smaller spiral galaxies
and the Local Group dwarf galaxies, where we expect larger dark matter
phase space densities and therefore shorter relaxation times.  Current
observations of the shapes of dark matter halos of nearby spiral
galaxies~\cite{Trachternach:2008wv} find, however, that their
gravitational potential is quite round, and that they seem to prefer a
central core. It is interesting to note that within our model, the
effect of dark sector Coulomb interactions could leave a dynamical
imprint in some of these nearby spiral galaxies but not the larger
elliptical galaxies. We leave more detailed investigations for future
work.

The case of the dwarf galaxies in the Local Group, that is, the
satellites of the Milky Way and Andromeda galaxies, is particularly
interesting. The stars in these galaxies have very small velocity
dispersions --- of the order of $10~\km/\s$ or even smaller. Combining
this with the observed extent of the stellar population, one may infer
the mass of the dark matter within the stellar extent of these dwarf
galaxies. The results of such an analysis show that the dwarf
satellites of the Milky Way are consistent with dark matter central
densities of about $0.1 M_\odot/\pc^3 \simeq
4~\gev/\cm^3$~\cite{Strigari:2008ib}. Unlike the galaxies we have been
considering previously, these dwarfs present a complication. Their
present day properties are set rather dramatically by the Milky Way
galaxy. As they fall into the Milky Way, gravitational tidal forces
will strip them of mass on the outside and thus reduce the velocity
dispersion of the dark matter particles inside the satellite's halo.
Exactly what the dispersion is depends on the initial mass, the extent
of the tidal mass loss and dark matter self-interactions.  If the
dispersion is of the order of $10~\km/\s$ (similar to the observed
stellar velocity dispersion), the energy transfer time scale is short
in the dwarf satellites for $\alpha_X$ and $m_X$ that are at the edge
of the allowed region in \figref{alpha_constraint}.  For these values
of $m_X$ and $\alpha_X$, interactions will almost certainly form cores
in the dwarf galaxies. Such cores are compatible with observations
unless they are larger than about 300
pc~\cite{Strigari:2006ue,Gilmore:2007fy}. Future astrometric
measurements of individual stars in these satellite galaxies will be
able to measure the density profile of dark matter halos on these
scales~\cite{Strigari:2007vn}.

Surprisingly, if we use the relation between the core radius and core
density for a self-gravitating isothermal sphere, $\rho_{\rm core} = 9
v_0^2/ (4 \pi G r_{\rm core}^2)$ with $v_0=10~\km/\s$ and $\rho_{\rm
core} \simeq 4~\gev/\cm^3$~\cite{Strigari:2008ib}, we obtain a core
size of 200 pc. However, our estimate has not accounted for some
crucial factors. If the mean free path is much shorter than the
typical orbit of the dark matter particles within the satellite, heat
conduction will be suppressed. In addition, the passage through the
Milky Way will introduce interactions with the dark matter in the
Milky Way and this could change the density profile of the satellite.

More detailed work is required to consider the effect of dark sector
Coulomb interactions on the satellites of the Milky Way and other
galaxies. We conjecture here that there would be regions of allowed
parameter space where the satellites galaxies and other small field
galaxies would show constant density cores in the center and reduced
substructure within their halos. This is beyond the scope of the
present work, but we urge the reader to keep in mind that future work
in this direction could lead to interesting astrophysical phenomena
and perhaps also rule out some of the allowed parameter space.

\section{Dark Matter in the Bullet Cluster}
\label{sec:bullet}

The Bullet Cluster is a rare system where a subcluster is seen to be
moving through a larger cluster. Through observations in the optical
and X-ray and strong and weak gravitational lensing observations,
astronomers have been able to map out the spatial distributions of the
stars, gas and dark matter in this system. From these inferred
distributions, it is clear that dark matter tracks the behavior of
stars, which are collisionless, rather than the gas. This
observation has allowed stringent bounds to be placed on the dark
matter self-interaction strength.

With respect to the self-interacting dark matter
proposal~\cite{Spergel:1999mh} discussed in the previous section
(velocity-independent cross section), the Bullet Cluster observations
demand~\cite{Markevitch:2003at,Randall:2007ph} that $\sigma_{\rm DM}
/m_X \alt 3000~\gev^{-3}$ ($\sigma_{\rm DM} /m_X \alt 0.7~\cm^2/\g$).
These are the most direct constraints on the self-interaction of dark
matter.

These bounds have been derived based on different considerations,
including the observed gas and dark matter offset, the high measured
subcluster velocity, and the survival of the subcluster after having
moved through the Bullet Cluster~\cite{Markevitch:2003at}. The last
phenomenon, survival of the subcluster, turns out to provide the
strongest constraint and hence we will focus on that.  We follow the
approach of Ref.~\cite{Markevitch:2003at} to derive bounds on the dark
sector Coulomb interaction, but relax the assumption of a hard-sphere
interaction cross section.

The subcluster experiences a net particle loss in a collision when the
velocities of both particles in the main and subcluster become larger
than the escape velocity $v_{\rm esc} \simeq 1200~\km/\s$.  This
condition may be turned into an effective range for the scattering
angle in the subcluster's reference frame
\begin{equation}
   \frac{v_{\rm esc}}{v_1} < \cos \theta <
   \left(1 - \frac{v^2_{\rm esc}}{v^2_1} \right)^{1/2} \ ,
\end{equation}
where $v_1 \sim 4800~\km/\s$ is the velocity of the main cluster
incoming particle before the collision.  We now assume that the
subcluster passed through the main cluster's center so that it saw a
surface density $\Sigma_m \sim 0.3~\g/\cm^2$. Demanding that the
fraction of particle loss be no greater than 30\%, we have an upper
bound on $\alpha_X$:
\begin{equation}
\label{bullet}
   f \equiv \frac{\Sigma_m}{m_X}\,
   \int d \Omega\, \frac{d \sigma}{d \Omega}
   = \frac{\Sigma_m}{m_X}\, \frac{2 \pi\, \alpha^2_X}{m^2_X\, v^4_1}\,
   \left[\frac{1}{1-\cos \theta_{\rm min}}
   - \frac{1}{1-\cos \theta_{\rm max}}
   \right] < 0.3\  .
\end{equation}
This bound is given in \figref{alpha_constraint}.  We see that it is
about four orders of magnitude weaker than that obtained from
considerations of the ellipticity of galactic dark matter halos.
Equation (\ref{bullet}) suggests that an improvement of this kind of
bound can be made by considering a galaxy falling into a cluster,
which results in a larger surface density $\Sigma_m = \int \rho_X dl
\sim 30~\g/\cm^2$, where the galaxy path is $l \sim 2 \pi R\, t_{\rm
age} / 6 \tau_{\rm dyn}$.  By demanding similarly that the galaxy does
not lose more than 30$\%$ of its particles during orbiting in the
cluster for about the age of the universe, we obtain a bound on
$\alpha_X$ roughly ten times stronger than from the Bullet Cluster, but
still weaker than the bounds obtained from the observed ellipticity of
galactic dark matter halos in the previous section.

\section{Conclusions}
\label{sec:conclusions}

We have investigated the astrophysical and cosmological consequences
of dark matter that is charged under an unbroken hidden $\uone$. We
find that this is a viable and natural dark matter candidate. The
salient arguments leading us to this conclusion are as follows.

\begin{itemize}
\item We investigated the dark matter-dark photon and dark matter-dark
  neutrino scattering processes in the early universe and found that
  the dark matter kinetically decouples early and therefore behaves as
  cold dark matter. The predictions for the temperature at which dark
  matter kinetically freezes out are shown in \figref{Tkd}.
\item Our calculations for the power spectrum of density
  fluctuations of dark matter showed that in these models (for $m_X >
  \gev$) we expect structure all the way down to at least $10^4
  M_\odot$. The predictions for these building blocks of structure in
  the Universe are summarized in \figref{Mc}. At the present time, we
  have no way to distinguish these hidden sector charged dark matter
  models from canonical WIMP models using matter power spectrum
  observations.
\item Although the dark matter particles in our model are cold, they
  also have long-range Coulomb interactions (but with a smaller fine
  structure constant for masses below 1 TeV). Since the hidden sector
  $\uone$ is unbroken, we have equal numbers of positively and
  negatively charged particles and hence we investigated the formation
  of bound states. We found that for masses below 10 TeV, bound state
  formation and the subsequent annihilation does not change the relic
  density of dark matter particles appreciably.
\item The self-interactions mediated by the long-range (hidden sector)
  Coulomb force can affect non-linear structure formation, especially
  at the small mass end. We found that the most stringent constraints
  arise from the observed ellipticity of galactic dark matter
  halos. (The self-interactions, if too strong, will make the core of
  dark matter halos round.)
\item Putting together all of the results above, we find that the mass
  of a dark matter particle charged under a hidden sector unbroken
  $\uone$ must be larger than about a GeV, {\em if} we restrict our
  attention to hidden sector weak mixing angles $\tantheta <
  10$. Larger values of the weak mixing angle will allow smaller dark
  matter particle masses. The detailed constraints imposed by the
  galactic dark matter halo observations, the Bullet Cluster
  observations and the requirement of obtaining the right relic
  density are shown in \figref{alpha_constraint}.
\item We also found that smaller galaxies such as the observed low
  surface brightness spirals and satellites of the Milky Way and
  Andromeda galaxy are likely to have constant density cores in their
  dark matter distribution. More work is required to pin down the core
  sizes, but this is clearly the regime where large differences from
  the predictions of collisionless cold dark matter, if any, will be
  found.
\end{itemize}

One of the promising avenues forward that we are currently exploring
is including a connector sector linking the hidden and visible
sectors. Such a connector sector will give rise to new phenomenology,
further constraints~\cite{Kamionkowski:2008gj,Carroll:2009dw}, and
interesting implications for a variety of dark matter detection
possibilities~\cite{Feng:2008dz,Feng:2008qn,Gardner:2008yn,%
McKeen:2009rm}.

A second promising avenue concerns structure formation on small
scales. Our formalism suggests that in the allowed parameter space,
the largest deviations from collisionless cold dark matter will appear
in galaxies much less massive than the Milky Way. A detailed
investigation of this aspect of hidden sector dark matter models is
beyond the scope of the present work, but simple estimates suggest
that, for large regions of model space, small galaxies will form cores
and the substructure in all dark matter halos will be reduced. These
expectations are in stark contrast to the predictions of collisionless
cold dark matter models.

We also stress that there are many other possibilities in the hidden
charged dark matter framework.  We have considered the hidden stau as
the dark matter.  Other charged particles, for example, the hidden
tau, chargino, quarks, and squarks, are also possible.  This scenario
also supports multi-component dark matter, with several hidden
sectors, each with its own dark matter particle contributing
significantly to the relic density through the WIMPless miracle.
Alternatively, even with only one hidden sector, stable hidden staus
and hidden (light, but massive) neutrinos could form mixed dark
matter, with both hot and cold components. All of these scenarios
merit further study.

To summarize, we have investigated the astrophysics and cosmology of
dark matter charged under a hidden unbroken $\uone$. We find that for
dark matter masses larger than about a GeV, these models have the
right relic density, make cosmological predictions currently
indistinguishable from the usual WIMP models, and are consistent with
observations on galactic and cluster scales.  At the same time, these
dark matter candidates are collisional and their annihilation is
enhanced, implying consequences for future observations and
experiments that may differ radically from WIMPs and other more
conventional possibilities.

\section*{Acknowledgments}

We thank Phil Humphrey, Jun Koda, Jason Kumar, and Quinn Minor for
many helpful discussions. The work of JLF and HT was supported in part
by NSF grant PHY--0653656. The work of MK was supported in part by NSF
grant PHY--0555689 and NASA grant NNX09AD09G.  The work of HY was
supported in part by NSF grants PHY--0653656 and PHY--0709742.

\appendix

\section{Calculation of Dark Matter Power Spectrum}
\label{sec:app}

In this appendix, we outline the calculation to compute the
small-scale power spectrum of dark matter.  In our hidden charged dark
matter scenario, the relevant interactions are elastic scattering off
both hidden sector photons and neutrinos.  For simplicity of
presentation, we present results below for just Compton scattering off
of photons.  This is the dominant process in most of parameter space.
In our numerical analysis and the results presented above, however, we
include also the neutrino scattering process.

The cross section for Compton scattering is
\begin{equation}
   \sigma_{\tilde{\tau}^h \gamma^h}=\frac{8\pi\alpha^2_X}{3m^2_X} \ .
\end{equation}
This differs markedly from the usual WIMP interaction cross section
with standard model fermions, which is suppressed relative to this by
$(T / m_\chi)^2$. This would result in lower decoupling temperatures
although the effect is mitigated to a large extent by the smaller
coupling required to obtain the correct relic abundance, as discussed
in \secref{kinetic}.

We study the Fourier-decomposed perturbations in the conformal
Newtonian gauge.  See Ref.~\cite{Ma:1995ey} for details.  For this
purpose it is better to use the visible photon temperature $T$ as the
``clock.''  Different hidden sector reheating temperatures $\ThRH =
\xiRH\, \TRH$ result in very different dark matter kinetic decoupling
temperatures $T_{\text{kd}}$ (cf. \figref{Tkd}). However, their effect
on the relation of the conformal time $\eta \equiv \int dt /a (t)$ to
the visible photon temperature is small.

For simplicity we assume that the fluctuation of the gravitational
potential is determined solely by the visible sector photons because
the energy density of the visible sector dominates.  With this
approximation we may use the analytic result for the Newtonian
potential in the radiation dominated regime,
\begin{equation}
   \Phi = 3 \Phi_P\, \left[\frac{\sin (k \eta /\sqrt{3}) -
   (k \eta /\sqrt{3})\,
   \cos (k \eta / \sqrt{3})}{(k \eta / \sqrt{3})^3} \right]\, ,
\end{equation}
where $k$ is the Fourier mode of interest and $\Phi_P$ is the
primordial amplitude. This approximation results in $\alt 10\%$ error,
depending on $\xiRH$.  The $\gamma^h$ perturbations may be expanded in
Legendre polynomials and this decomposition results in a multipole
hierarchy:
\begin{eqnarray}
   \nonumber\dot{\Theta}^h_0+k\Theta^h_1&=&-\dot{\Phi}\, \nonumber \\
   \dot{\Theta}^h_1+k\left(\frac{2}{3}\Theta^h_2-\frac{1}{3}\Theta^h_0\right)
   &=&\frac{k\Psi}{3}-\dot{\tau}\left(\Theta^h_1-\frac{\theta_X}{3k}\right)
   \nonumber \\
   \dot{\Theta}^h_2-\frac{2k}{5}\Theta^h_{1}
   &=&-\frac{9}{10}\dot{\tau}\Theta^h_2\, \nonumber \\
   \dot{\Theta}^h_l &=& \frac{k}{2l +1}\, \left[l\, \Theta^h_{l-1}
   - (l+1)\, \Theta^h_{l+1} \right]\, - \dot{\tau}\, \Theta^h_l\,
\hspace{0.3cm}
   {\rm for} \hspace{0.2cm} l \geq 3\, ,
\end{eqnarray}
where the $\Theta^h_l$ are the multipole moments of the hidden sector
temperature field, to a good approximation $\Psi = - \Phi$,
$\dot{\tau} \equiv a\, n_X\, \sigma_{\tilde{\tau}^h \gamma^h}$ is the
scattering rate, and $n_X$ is the $\tilde{\tau}^h$ number density.
The dark matter density and velocity perturbations are
\begin{eqnarray}
  \nonumber\dot{\delta}_X + \theta_X &=& -3\dot{\Phi} \nonumber \\
  \dot{\theta}_X + \frac{\dot{a}}{a}\, \theta_X
  &=& - k^2 \Phi + k^2 c^2_s\, \delta_X - k^2 \sigma_X +
  \frac{3\dot{\tau}}{R}
\left( k \Theta^h_1 - \frac{\theta_X}{3} \right)\, ,
\end{eqnarray}
where $R\equiv3\rho_X/4\rho_{\gamma^h}$ is the ratio of dark matter to
hidden sector photon energy density.  The interaction term
\begin{eqnarray}
   \frac{3\dot{\tau}}{R}=\frac{32\pi^3\alpha^2_X}{45}
   \frac{T^{h\, 4}}{m^3_X}a
\end{eqnarray}
is the rate of transferring momentum $|\vec{p}| \sim T^h$ from the
hidden sector photons to the dark matter particles (cf. the second
term in \eqref{rate}).  We follow the treatment in
Ref.~\cite{Loeb:2005pm} to set the dark matter sound speed $c_s$ and
shear $\sigma_X$.

The main difference between our hidden sector scenario and standard
cosmology is that the hidden sector photon decoupling epoch is much
earlier than the drag epoch of the hidden charged dark matter.  We
include hidden sector photon higher multipoles and truncate the series
at $l_{\rm max} = 10$ to accurately account for the free-streaming of
photons.

To obtain the transfer function we solve the dark matter fluid
equation well into the decoupling regime and the evolve further in
time (when the calculation becomes computationally intensive) using
the free-streaming solution~\cite{Loeb:2005pm}
\begin{eqnarray}
  \delta_X(\eta)=\exp\left[-\frac{1}{2}\frac{k^2}{k^2_f}\ln^2
\left(\frac{\eta} {\eta_\ast} \right)
\right]\left[\left. \delta_X \right|_{\eta_\ast}+
\left. \frac{d\delta_X}{d\eta}\right|_
  {\eta_\ast}\eta_{\ast}\ln \left(\frac{\eta}{\eta_\ast} \right)\right],
\end{eqnarray}
where $k^{-2}_f= \eta^2_\ast T^h(\eta_\ast) / m_X$, to the time of
matter-radiation equality.



\end{document}